# Demonstration of ultrafast coherent sub-picosecond magnon pulses

P. Kubaščík[1*], M. Leiviskä[2], M. Bušina[1], Z. Kašpar[1,2], K. Müller[3,4], J. Mak[5], M. Scheufele[3,4], G. Lemut[5], J. Solano[2], O. Franke[5], R. Matuška[2], J. Jechumtál[1], E. Schmorranzerová[1], P. Němec[1], P. Kužel[2], T. Kampfrath[5], P. W. Brouwer[5], M. Opel[3], S. Geprägs[3], M. Althammer[3,4], T. Jungwirth[2,6,7], H. Reichlova[2**], L. Nádvorník[1***]

1. Faculty of Mathematics and Physics, Charles University, 121 16 Prague, Czech Republic
2. Institute of Physics, Czech Academy of Sciences, Prague, Czech Republic
3. Walther-Meissner-Institut, Bayerische Akademie der Wissenschaften, Garching, Germany
4. Technical University of Munich, TUM School of Natural Sciences, Physics Department, Garching, Germany
5. Department of Physics, Freie Universität Berlin, 14195 Berlin, Germany
6. School of Physics and Astronomy, University of Nottingham, Nottingham, UK
7. Center for Science and Innovation in Spintronics, Tohoku University, Sendai, Japan

* E-mail: peter.kubascik@matfyz.cuni.cz

** reichlh@fzu.cz

*** lukas.nadvornik@matfyz.cuni.cz

## Abstract

Ultrafast magnonics explores routes towards energy-efficient information processing by replacing electrons with pulses of terahertz magnons. The critical, but so far elusive, requirement on the magnon pulses is the combination of high-speed and a long-range propagation while maintaining a well-defined narrow waveform of the pulses. Here, we demonstrate the launching of 0.4-picosecond magnon pulses in the altermagnetic insulator $\alpha$-$Fe_2O_3$ (hematite) using broadband terahertz spin-orbit torques by an adjacent platinum layer. Phase-resolved terahertz transmission reveals that these sub-picosecond pulses travel ballistically, coherently and non-dispersively at a constant velocity of 23 nm/ps, transport angular momentum and retain their temporal waveform shape over 160 nm, with an inferred relaxation length in hundreds of nanometers or more. Beyond establishing $\alpha$-$Fe_2O_3$ as an optimal platform for ultrafast magnonics, our technique introduces a high-resolution, tabletop method to directly extract dispersion relations up to high wavevectors, and explore ultrafast magnon dynamics in altermagnets and other magnetic materials with teraherz-range dynamics.

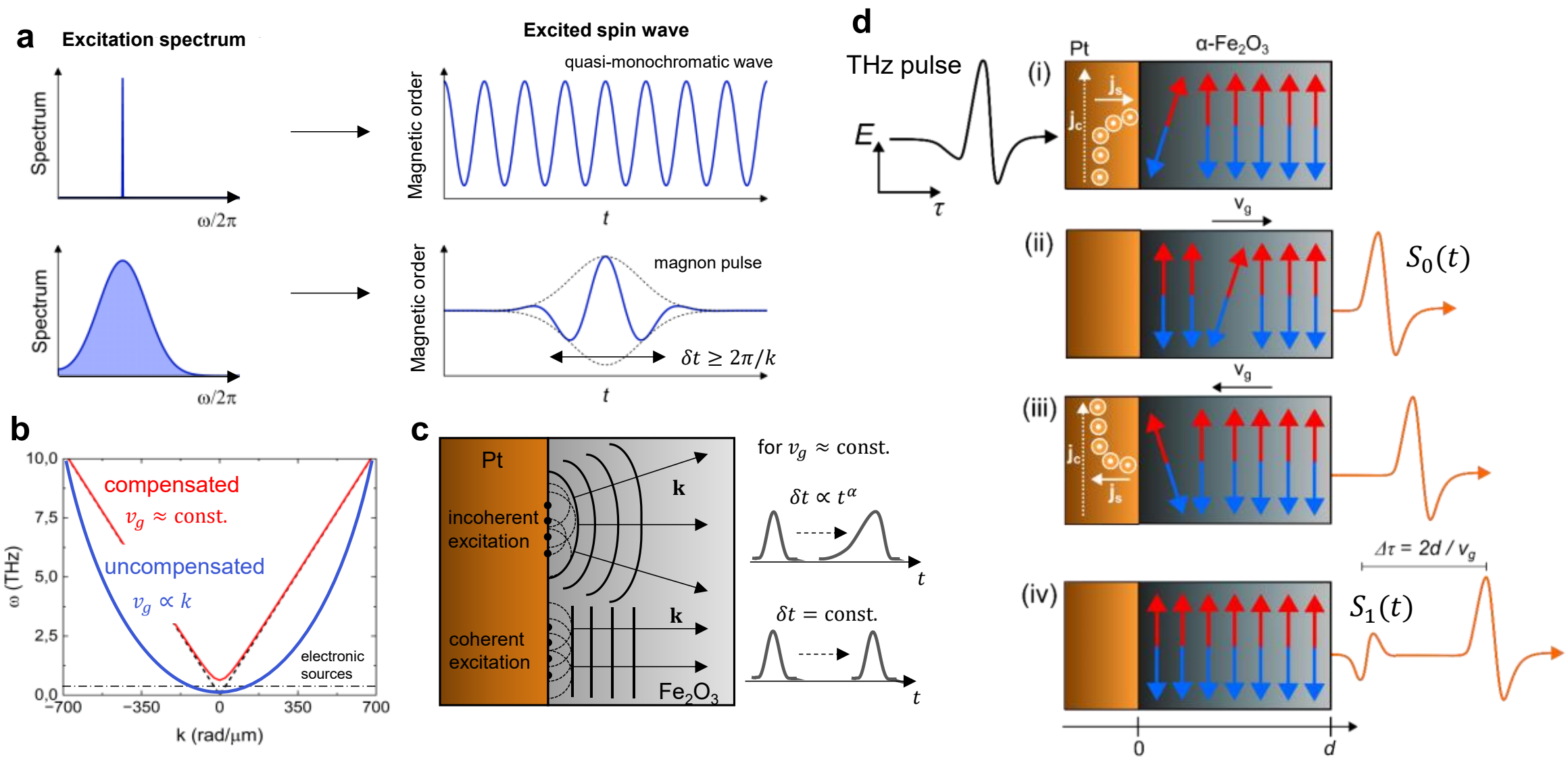

**Fig. 1 | Excitation of magnonic pulses. a**, An illustration of the requirement on broadband excitation for launching short magnon pulses. A narrowband, monochromatic coherent excitation leads to long-lasting plane-wave-like oscillations, incompatible with high clock-speed operation. On the contrary, a broadband coherent excitation is needed for short pulse creation. **b**, Magnonic dispersion for compensated (red) and uncompensated magnets (blue), highlighting the constant and $\boldsymbol{k}$-dependent group velocity and consequent pulse broadening. The frequency region accessible by conventional electronics is shown by horizontal dashed line. **c**, An illustration of geometrical broadening of pulses. The incoherent, thermal excitation does not maintain the global phase in the excitation region and the wavefront is parabolic, resulting in all-angle propagation and consequent pulse broadening. The coherent excitation keeps the global phase locked, forming a straight wavefront and propagation solely perpendicular to the interface with no geometrical broadening. **d**, Principle of magnon launching: The THz pulse induces the in-plane charge current $\boldsymbol{j_c}$ and, due to the spin Hall effect, the perpendicular spin current $\boldsymbol{j_s}$ and spin accumulation at the Pt/$Fe_2O_3$ interface. Here, it launches a magnon, fingerprinted in modified THz transmission $\boldsymbol{S_0(t)}$. Simultaneously, the propagating magnon pulse reflects from the other boundary of the $Fe_2O_3$ layer, returns to the interface and radiates an additional THz pulse $\boldsymbol{S_1(t)}$ via the inverse mechanism.

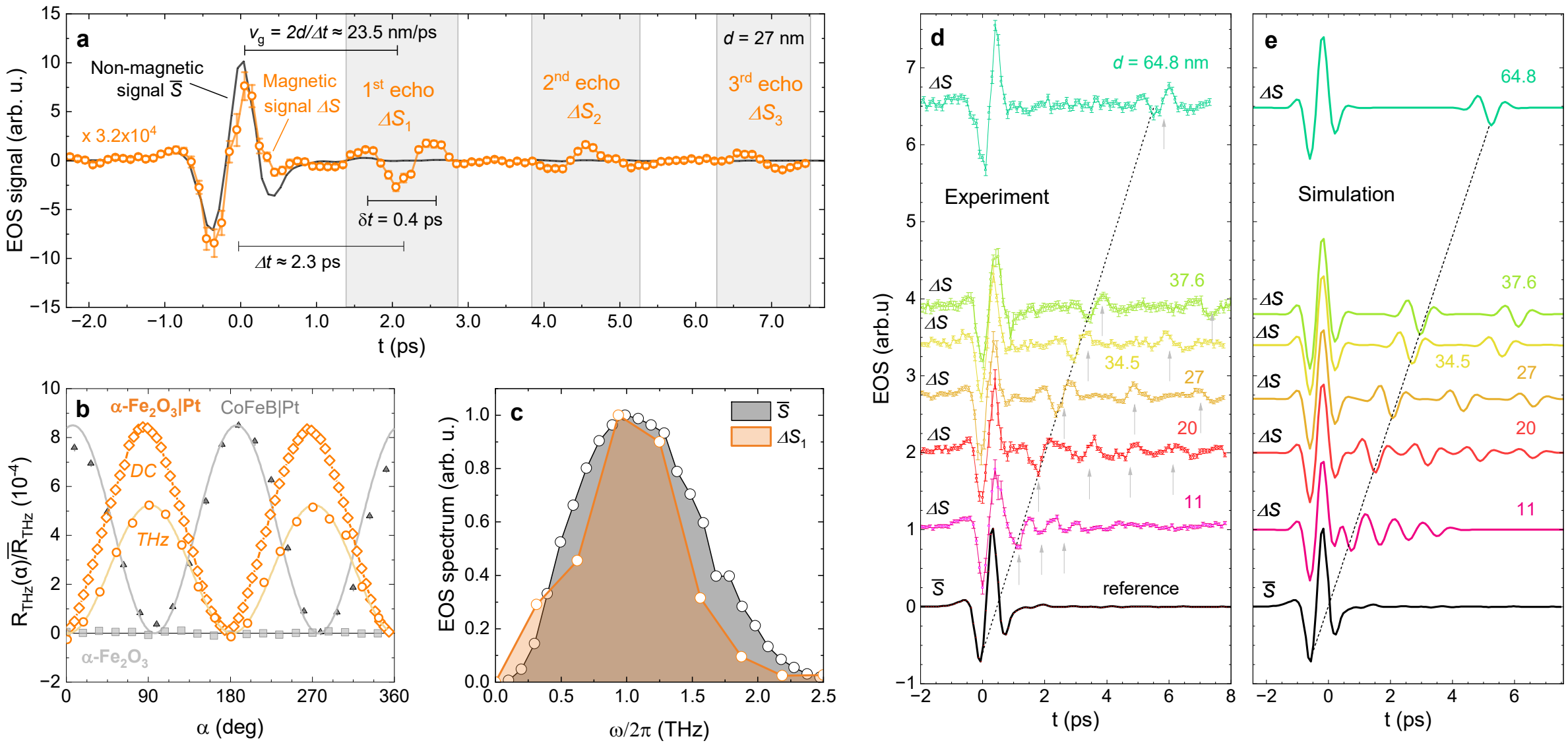

**Fig. 2 | Ultrashort magnon pulses. a**, THz response of excitation of Pt|α-Fe2O3($d$), $d = 27$ nm, revealing three magnon pulse reflections $\Delta S_i$ following the SMR transmission trace $\Delta S_0$ in the magnetic signal $\Delta S$ (orange data points), contrasted to the non-magnetic signal $\bar{S}$ (black curve). **b**, The THz SMR as a function of angle $\alpha$ between the direction of the magnetic order and polarization of THz pulse for Pt| $\alpha$-$Fe_2O_3$(27 nm) (orange circles) and for $\alpha$-$Fe_2O_3$(27 nm) only (grey squares), complemented by the DC experiments on Pt| $\alpha$-$Fe_2O_3$(27 nm) (orange diamonds) and on the ferromagnetic reference Pt|CoFeB (gray triangles). Curves are fits by a phase-shifted $cos(2\alpha)$ function. **c**, The spectrum of $\bar{S}(t)$ and $\Delta S_1(t)$ with the maximum normalized to unity for the same sample. **d**, $\Delta S(t)$ for samples with $d$ ranging from 11 to 65 nm and reference $\bar{S}(t)$. The linear trend of delays of first magnon reflections with $d$ is highlighted by the dashed line, all magnon traces are indicated by the arrows. **e**, The output of the transport model for fitted parameter $v_g = (23 \pm 1)$ nm/ps. In **a**, **d** and **e**, the $\Delta S(t)$ is multiplied by a factor of $2.3 \times 10^4$ for comparison.

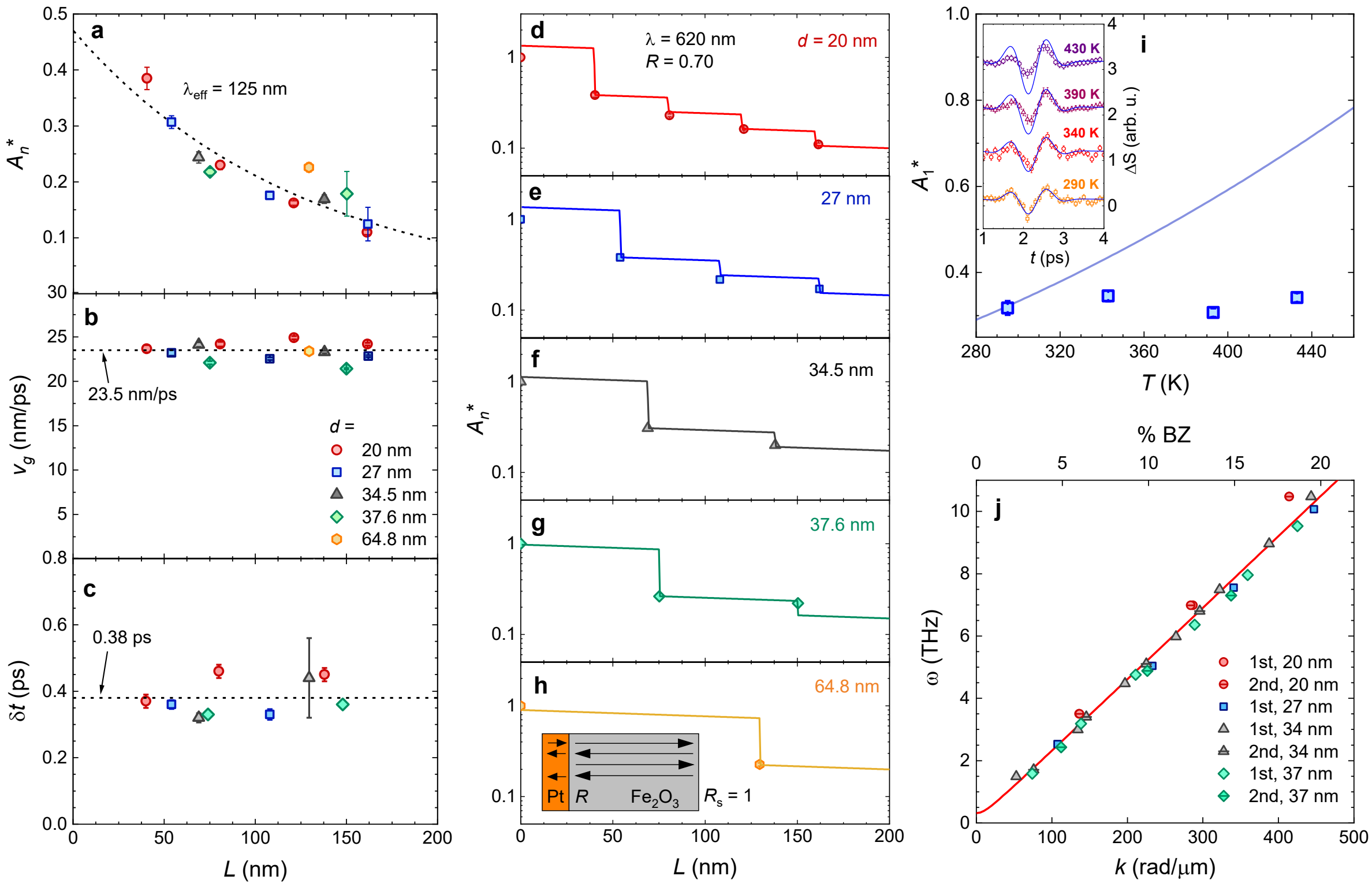

**Fig. 3 | Parameters of the magnon transport. a**, The rms amplitude $A_{ni}$ of $\Delta S_i(t)$ for all samples, normalized to rms of respective $\Delta S_0(t)$. The empirical single-exponential fit yields $1/e$ travelled distance of 125 nm. **b, c**, the group velocities $v_g$ and durations $\delta t$ inferred for various magnon pulses, fitted by a constant (dashed line). **d–h**, $A_n^*$ as a function of $L$ for each $d$ compared to a universal fit by the leaky resonator model for coherent (thick curves) and incoherent magnons (dashed curves). Inset: an illustration of the leaky resonator model for magnon internal reflections in the $\alpha$-$Fe_2O_3$ layer. **i**, the rms of $\Delta S_1(t)$ for $d = 27$ nm as a function of sample temperature $T$, compared to the expectation from a model for incoherent magnons (blue curve). Inset: the respective $\Delta S_1(t)$ traces. The blue curve is a replica of $\Delta S_1(t)$ for $T = 290$ K for comparison. **j**, The inferred dispersion $\omega(k)$ from $\Delta S_1(\omega)$ spectra for several samples. The red line is the model of magnon dispersion in antiferromagnets.

Magnons offer a promising platform for charge-free and ultrafast information processing, but building high-speed all-magnon transistors [1] and phase-sensitive logic devices [2] requires well-defined, sub-picosecond pulses to achieve terahertz clock rates. Launching and transporting such ultrashort pulses over long distances at high velocities without distortion imposes four simultaneous requirements [3], as illustrated in Fig. 1. First, a broadband and sufficiently high-frequency excitation is needed for a sub-picosecond temporal envelope which favors impulsive THz excitation rather than resonant narrowband driving (Fig. 1a). Second, magnon bands with linear dependence of frequency on wavevector, inherent to exchange-dominated compensated magnets, such as antiferromagnets or altermagnets, is crucial to minimize wavepacket chirping due to group-velocity dispersion (Fig. 1b). Third, coherent coupling to the excitation impulse is required to preserve the well-defined phase of magnons and prevent their geometric broadening due to multi-angle propagation [4,5,6], typical for incoherent, thermally driven magnons (Fig 1c). Finally, the magnon transport needs to be of ballistic, scattering-free nature to allow for long-distance, high-speed and broadening-free propagation.

Despite considerable effort, no experiment has reported a magnon excitation that satisfies all of these conditions simultaneously [3]. THz magnetic-field [7,8], non-uniform light absorption [9], impulsive stimulated Raman scattering [10], ultrafast spin-orbit-torque [11,12], radio-frequency [13,14] and strain-mediated [15] excitation schemes have demonstrated coherent magnon generation in ferromagnetic, antiferromagnetic and altermagnetic insulators. However, all these attempts invariably drove either narrowband or resonant, i.e. harmonic plane-wave-like, dynamics with wavevector $k \approx 0$ and no transport properties reported [7,16,17,18,19,20], or single or multi-mode transport with higher $k$-vectors and reported velocities but also wide, tens or hundreds of picosecond-long envelops and decays [8,9,21,22]. Finally, standing high-$k$ modes rather than propagating pulses have been also reported in ferromagnets [23]. Conversely, when well-defined pico- and sub-picosecond propagating pulses have been demonstrated, however, the transported magnons were of incoherent origin, therefore they suffer from geometrical broadening and reach only a few tens of nanometers before vanishing [4]. In all these cases, the excitation of magnetic order either fails to form a well-defined ultrashort envelope, i.e. does not meet the requirement for high clock-speed magnon devices, or the formed pulse broadens rapidly due to geometrical broadening and relaxes before propagating over a useful distance. A magnon excitation and propagation that is simultaneously sub-picosecond, coherent, non-dispersive, and ballistic over hundreds of nanometers has, so far, been an elusive goal of ultrafast magnonic research.

Here, we report this unique regime of magnonic transport by means of a table-top technique. Using broadband THz excitation of the altermagnetic insulator $\boldsymbol{\alpha}$-$Fe_2O_3$ (hematite) and phase-resolved THz transmission detection, we observe coherent magnon pulses with a duration of 0.4 ps that transport angular momentum and propagate ballistically over 160 nm while retaining their temporal shape, i.e. forming a non-dispersive pulse. The pulses travel ballistically at a constant velocity of 23 nm/ps, in quantitative agreement with the group velocity expected for the linear-in-momentum part of the magnon band structure of $\boldsymbol{\alpha}$-$Fe_2O_3$, and with the inferred magnon relaxation length in the order of hundreds of nanometers or more in the absence of interfaces. This is an order of magnitude longer than the relaxation lengths reported for incoherent magnon transport [4]. These results demonstrate that all four requirements for a pulse compatible with ultrafast magnon logic, i.e. broadband coherent excitation and ultrafast, non-dispersive and long-range ballistic propagation, can be met simultaneously in $\boldsymbol{\alpha}$-$Fe_2O_3$, establishing this readily available high-temperature altermagnetic material as a platform for ultrafast magnon logic.

## Launching of a magnon pulse

The magnon launching principle is illustrated in Fig. 1d. A picosecond THz pulse is absorbed in a 3-nm-thin platinum cap of a $\boldsymbol{\alpha}$-$Fe_2O_3$ layer of various thicknesses $d$. The electric field of the THz pulse, oriented at an angle $\alpha$ relative to the magnetic order of $\boldsymbol{\alpha}$-$Fe_2O_3$, drives an in-plane electric current $j_c$ and, via the spin Hall effect, induces a transient electronic spin accumulation at the interface. The transient angular momentum couples to the magnetic moments in $\boldsymbol{\alpha}$-$Fe_2O_3$ and triggers magnon excitations via the coherent spin torque. This phenomenon has been established as a mechanism of the spin Hall magnetoresistance (SMR) [24] and has been also very recently reported in the THz frequency range in ferrimagnetic yttrium iron garnet [25]. The SMR effect itself is manifested by a $cos(2\alpha)$-dependent variation of the electrical resistance and, thus, of the transmitted THz pulse $S_0(t,\alpha)$. If the aforementioned conditions are met, the triggered magnon excitation propagates as a wavepacket through the $\boldsymbol{\alpha}$-$Fe_2O_3$ layer, is reflected on the $\boldsymbol{\alpha}$-$Fe_2O_3$ -substrate interface and arrives back at the platinum interface and is partially transmitted into the platinum layer. Here, it is converted via the inverse spin Hall mechanism and radiates another THz pulse $S_1(t,\alpha)$ with a temporal delay $\Delta t$ after the initial pulse. This mechanism strictly requires that the magnon pulse transports angular momentum. In the case of easy-plane $\boldsymbol{\alpha}$-$Fe_2O_3$ it implies a chirality of the mode [26,27].

A typical example of the time-dependent THz responses of our Pt|$Fe_2O_3$ sample with $d = 27\ \mathrm{nm}$ is shown in Fig. 2a. The magnetic signal $\Delta S(t)$ (orange data points), which is given by the difference of THz signals $S(t,\alpha)$ for $\alpha = 0°$ and $90°$, shows the initial SMR pulse $\Delta S_0(t)$, followed by three echoes, $\Delta S_n(t),\ n = 1,2,3$, with pulse duration $\sim 0.4\ \mathrm{ps}$, evenly separated by ~ 2.3 ps. By assuming the above-described mechanism with multiple reflections, the approximated velocity of propagation is $v_{\mathrm{g}} = 2d/\Delta t \approx 24\ \mathrm{nm/ps}$, consistent the high-$k$ group velocity reported in $\boldsymbol{\alpha}$-$Fe_2O_3$ [28]. To confirm the origin of the echoes in the travelling magnon pulses, we first compare the signal with the non-magnetic THz transmission signal $\bar{S}(t)$, given by the sum of signals for $\alpha = 0°$ and $90°$ (black curve in Fig. 2a). It shows the driving THz pulse, but it lacks any signatures of the echoes. Furthermore, the $\cos(2\alpha)$ symmetry of the maximum of $\Delta S_0(t)$ vs $\alpha$ has been confirmed in Fig. 2b, where the signal was recalculated into variation of electrical resistance $R(\alpha)$ (see Methods). The same experiment performed on $Fe_2O_3$ without the platinum cap shows no scalable signal and corroborates the SMR as the underlying mechanism of the $S_n$ pulses. The DC experiment with the same symmetry and amplitude by a factor of two as compared to the THz one is consistent with literature [29,30] and with the decay of the SMR signal with frequency in insulating systems [25]. The SMR signals on Pt|$Fe_2O_3$ also show the 90° phase shift in $\alpha$ as compared to the analogous measurement on the reference ferromagnetic Pt|CoFeB sample. This is a characteristic feature of the SMR on antiferromagnets where the magnetic order is manipulated by the small uncompensated canted moment due to Dzyaloshinskii-Moriya interaction [30,31]. The spectral composition of the pulse $\Delta S_1(t)$ (Fig. 2c) indicate similar structure as the driving force represented by $\bar{S}(t)$, indicating the coherent excitation regime.

The ultimate evidence for the magnon-pulse nature of the echoes is obtained when the thickness of the $\boldsymbol{\alpha}$-$Fe_2O_3$ layer is varied from $d = 11$ to $65\ \mathrm{nm}$, as shown in Fig. 2d. The dashed line highlights the linear trend of increasing delays with travelled distance, $L = 2d$, of the first echo in $\Delta S(t)$ from the temporal position of the main driving pulse $\bar{S}(t)$, confirming the propagation character of the magnon wave packets. The observed dynamics is also compared to a modeling based on Landau-Liftshitz-Gilbert and Boltzmann transport formalism, shown in Fig. 2e, for fitted $v_g = (23 \pm 1)\ \mathrm{nm/ps}$ (other parameters and model details are summarized in the Methods section). The varying shape of echoes is attributed to the small deviations from the strict linearity of the magnon dispersion and the effect of reflection at the interfaces.

## Characteristics of the magnon transport

More insight into the magnon transport is provided in Fig. 3. The decay of normalized root-mean-square (rms) amplitudes $A_n^*$ of the $n$-th magnonic echo $\Delta S_n$ in all studied samples, normalized to the rms $A_0$ of the main SMR response $\Delta S_0$, is shown in Fig. 3a. The decay of $A_n^*$ is due to the radiative outcoupling from the magnonic resonator and magnon relaxation. First, without detailed modelling, $A_n^*(L)$ can be empirically fitted by a single exponential, yielding an $1/e$ effective transport length $\lambda_{\mathrm{eff}} = (125 \pm 18)\ \mathrm{nm}$. Considering the delayed arrival $\Delta t$ of echoes, calculated as the time-weighted average of $\Delta S_i(t)$ [4] (see Methods), the extracted group velocities $v_g$ are approximately independent of $L$ with the magnon mean free path of at least 160 nm, and mean value of $v_g = (23 \pm 1)\ \mathrm{nm/ps}$ (Fig. 3b) is in striking match with the high-$k$ literature values [28]. The observed $L$-independent $v_g$ proves that the transport is predominantly ballistic, since the diffusive transport regime would be fingerprinted by the $v_g \propto L^{-1}$ dependence [4]. The ballistic transport regime also implies that the mean time between magnon scattering events must be significantly longer than the time needed to travel the largest $L$ observed in Fig. 3b, i.e., $\tau \gg L/v_g \approx 7\,\mathrm{ps}$. Similarly, we observe no broadening of the magnon pulses as their temporal width $\delta t = (0.38 \pm 0.06)\,\mathrm{ps}$ is also maintained constant with $L$, in contrast to $L^2$ dependence expected for the diffusive regime (Fig. 3c). The $L$-independence of $\delta t$ further implies that there is no observable geometrical broadening, that would be typical for incoherent magnons. In that case, the global phase in the excitation area is not maintained, forming a non-planar wave front and causing magnons propagate under a distribution of angles (Fig. 1c). This naturally leads to a distribution of arrival times at the platinum interface and causing additional broadening and smearing of the magnon pulse [4,5,6,32].

The achievable magnon transport length in this regime is addressed in Figs. 3d-h. Here, we use the leaky resonator model (inset in Fig. 3h and Methods), in which the echoe amplitudes are described by $A_n^* = \gamma(1-R)R^{n-1}e^{-2nd/\lambda}$, where $\gamma$ is a normalization factor set by fitting the value of $A_1^*$ as a boundary condition, $\lambda$ is the magnon relaxation length in absence of interfaces and $R$ is the magnon reflection coefficient for the Pt|$\alpha$-$Fe_2O_3$ interface ($R_s = 1$ is considered for the $\alpha$-$Fe_2O_3$|substrate interface). Within this description, $\lambda > \lambda_{\mathrm{eff}} \approx 125\,\mathrm{nm}$, which is the lower bound for the magnon transport length scale. The optimal fits of $A_n^*$ by this model yield $R \approx 0.70$ and $\lambda \approx 620\,\mathrm{nm}$ universal for all samples (see Methods for details), implying the magnon relaxation time, $\tau_{\mathrm{rel}} \sim 25\,\mathrm{ps}$. However, the model provides an acceptable match also for $R \approx 0.6$ with $\lambda$ reaching micrometer distances and $\tau_{\mathrm{rel}}$ tens of picosecond. Interestingly, $\tau_{\mathrm{rel}}$ represents here either the life time of the magnon itself, or the relaxation time of its chiral mode. The model for coherent magnons can also be extended to $L = 0$ and compared to the SMR transmission response $A_0$. From the slight overestimate of the fit at $L = 0$, best visible for the thinner samples, we can infer that $A_0$ contains also incoherent components that effectively lead to a decrease of the SMR signal [25]. However, as the incoherent magnons do not form a non-dispersive travelling pulse, they do not contribute to the echoes.

The coherent nature of travelling magnon pulses is corroborated in Fig. 3i which shows $A_1^*$ for room temperature and elevated temperatures $T$. For $T$ far below the Néel temperature, the thermal activation of incoherent magnons would imply a $T^2$ dependence. However, we observe an approximately $T$-independent $A_1^*$. The natural phase sensitivity and large bandwidth of our method allows us to extract broadband phase change $\Delta\phi(\omega) = \arg[S_1(\omega)/S_0(\omega)]$ from the first magnon echoes and access the dispersion $\omega(k)$ of our magnon pulses by inverting $k(\omega) = \Delta\phi(\omega)/L$, as shown in Fig. 3j [33]. Over a large span of $k$-vectors, with the maximum around 470 rad/$\mu m$, we verified the expected linear dispersion [28], and demonstrated table-top probing of approximately 20% of the Brillouin zone.

## Outlook

Our observations establish the coherent, broadband terahertz excitation of compensated magnets with an adjacent heavy-metal layer as a direct pathway to generate ultrashort, angular-momentum-transporting magnon pulses that satisfy essential prerequisites for high-clock-speed magnon logic. The purely ballistic transport of sub-picosecond pulses without measurable geometrical or group-velocity broadening over distances of more than hundred nanometers and with inferred magnon relaxation length of several hundreds of nanometers, an order of magnitude longer than previously reported for incoherent magnons, overcomes a critical bottleneck in magnonics. The robust generation and detection of these non-dispersive pulses in $\alpha$-$Fe_2O_3$ not only validates $\boldsymbol{\alpha}$-$Fe_2O_3$ as a prime candidate for future ultrafast magnonic circuitry but also establishes the SMR in the THz regime as a method for efficient transduction between the charge and magnon domains.

Beyond device applications, our methodology provides a uniquely sensitive, phase-resolved probe of magnetic dynamics at the nanoscale. The ability to directly access the dispersion relations at large $k$-vectors using a magnon-chirality sensitive tabletop technique opens new possibilities for investigating fundamental magnon properties. This includes exploring the highly anisotropic magnon spectra with alternating chirality splitting in altermagnets, where $\boldsymbol{\alpha}$-$Fe_2O_3$ itself has recently been recognized as a prominent material example and mapping non-reciprocal or topological magnon features in other unconventional magnetic materials. Our work thus bridges fundamental research of magnon excitations with the oriented research towards ultrafast magnonic information technologies.

# Methods

## Samples and growth

The α-$Fe_2O_3$|Pt samples were grown by pulsed laser deposition of α-$Fe_2O_3$ on (0001)-oriented single-crystalline sapphire substrates, followed by in situ deposition of platinum by electron-beam evaporation. The resulting stack is sapphire|α-$Fe_2O_3(d)$|Pt($d_{\mathrm{HM}}$), with $d = 11, 20, 27, 34.5, 37.6$ and $64.8\ \mathrm{nm}$ and $d_{\mathrm{HM}} = 0$ and 3 nm, where $d_{\mathrm{HM}} = 0$ is a control sample.

## DC characterisation

The α-$Fe_2O_3$|Pt thin films were patterned into Hall bar structures and the longitudinal four-point resistance was measured using the Delta method. The sample was mounted in a vector magnet and a magnetic field of 0.3 T was applied in the measurements presented.

## THz experiments

Pulses of THz radiation are generated by optical rectification of femtosecond laser pulses from a regenerative amplifier (wavelength 1030 nm, duration 170 fs, energy 50 µJ, repetition rate 10 kHz) in a GaP(110) crystal (thickness 2 mm). After transmission through the sample, the THz electric field is characterized by electro-optic sampling (EOS) in GaP(110) (2 mm). The external magnetic field is provided by a rotatable Halbach array with an amplitude of ≈ 350 mT. The magnetic signal $\Delta S$ was measured in crossed polarizer geometry, when the sample was placed inside of the magnet and surrounded by two polarisers. First, the polarizers were rotated parallelly and standard, nonmagnetic, EOS signal transmitted through the sample was measured [$S_{\parallel}(t)$]. Next, the polarires were crossed and transmitted EOS signal was measured [$S_{\perp}(t)$] was measured for different orientations of magnetic field.

The $S_{\parallel}(t,\alpha)$ and $S_{\perp}(t,\alpha)$ was then converted to frequency space [$S_{\parallel}(\omega,\alpha)$ and $S_{\perp}(\omega,\alpha)$] and the change in off-diagonal resistivity (Fig. 2b) was then calculated using [34,35]

$$\frac{R_{xy}(\alpha,\omega)}{\bar{R}(\omega)} = \frac{\mathrm{S}_{\perp}(\alpha,\omega)-\bar{\mathrm{S}}_{\perp}(\omega)}{S_{\parallel}(\omega)}\left[1+\frac{n(\omega)+1}{Z_0\,\bar{R}(\omega)}\right],$$

where $\alpha$ is the angle between THz polarisation and applied magnetic field and $\bar{\mathrm{S}}_{\perp}(\omega)$ is average of $\mathrm{S}_{\perp}(\alpha,\omega)$ over all directions of magnetic field.

Because of SMR, the symmetry of $\mathrm{S}_{\perp}(\alpha,t)$ follows

$$S_{\perp}(\alpha,t) = \bar{\mathrm{S}}_{\perp}(t) + \Delta S(t)\sin(2\alpha) \quad (1)$$

dependence. The magnetic signal $\Delta S(t)$ was obtained by fitting $S_{\perp}(\alpha,t)$ by Eq. 1 for each $t$.

## Quantification of magnon echoes

Amplitude of $n$-th magnon echo $A_n^*$ was estimated as RMS value of $\Delta S(t)$ signal over window of 3.1 ps and normalized by signal around the main pulse,

$$A_n^* = \frac{RMS(\Delta \mathrm{S}_n(t))}{RMS(\Delta \mathrm{S}_0(t))}$$

The group velocity was estimated by

$$v_g = 2nd/(t_n - t_0),$$

where $t_n$ is the average time delay of $n$-th echo calculated as weight average of $\Delta S(t)$ in time domain,

$$t_n = \frac{\int |\Delta S_n(t)| t \mathrm{d}t}{\int |\Delta S_n(t)| \mathrm{d}t},$$

where $|\Delta S_n(t)|$ is the weight function. The temporal width $\delta t$ of the echoe was consequently calculated as weighted standard deviation of time delay around the $t_n$

$$\delta t_n = \sqrt{\frac{\int |\Delta S_n(t)| (t - t_n)^2 \mathrm{d}t}{\int |\Delta S_n(t)| \mathrm{d}t}},$$

where also$|\Delta S_n(t)|$ is the weight function.

The uncertainties assigned to individual values of $t_n, A_n^*$ and $\delta t_n$ were estimated using a modified jackknife procedure [36]. Each data point was omitted in turn, and the quantity was recalculated from the remaining points. The average values of group velocity $v_\mathrm{g}$ and temporal width $\delta t$ was estimated as average of the group velocities of all echoes and its uncertainty as their standard deviation, representing the variation between the samples. The uncertainty of $\lambda_\mathrm{eff}$ was estimated as a uncertainty of regression coefficient after direct fitting by single exponentail decay.

**Data availability**

**Acknowledgments**

This work was supported by TERAFIT project No. CZ.02.01.01/00/22_008/0004594 funded by Ministry of Education Youth and Sports of the Czech Republic (MEYS CR), programme Johannes Amos Comenius (OP JAK), call Excellent Research. The authors acknowledge funding by the Czech Science Foundation (grant no. 21–28876J), by the Grant Agency of the Charles University (grants no. 166123, no. 120324 and SVV–2025–260836), by CzechNanoLab Research Infrastructure supported by MEYS CR (LM2023051), by MEYS CR project LNSM-LNSpin, by the German Research Foundation through the collaborative research center CRC/TRR 227 “Ultrafast spin dynamics” (project ID 328545488, projects A05, B02 and B03), and the Research Unit RU 5844 ChiPS (Project No. 541503763). K.M. and M.A. acknowledge funding by the European Union's Horizon Europe research and innovation program under Grant No. 101171325 (POSA). The authors acknowledge the networking opportunities provided by the European COST Action No. CA23136 (CHIROMAG).

---